# Reconstruction of the Tunguska Event of 1908: Neither an Asteroid, Nor a Comet Core

Vladimir Rubtsov

International Institute of Environmentally Safe Technologies

P.O. Box 4542, 61022 Kharkov-22, Ukraine

E-mail: Tolimak@gmail.com

*Abstract:* The Tunguska explosion occurred in the morning of June 30, 1908, in Central Siberia, some 800 km NNW from Lake Baikal. It devastated the forested area of 2150 km$^2$, flattening and scorching some 30 million trees. Before this, a luminous body flew overhead in the cloudless sky. The air waves from the explosion were recorded as far as in London. The object that flew that morning over Siberia is usually designated the "Tunguska meteorite" or – more cautiously – the "Tunguska space body" (TSB). Certainly, this body was dangerous: the taiga was leveled over an area twice as large as New York City. The whole number of Tunguska hypotheses reaches a hundred, or so. But few of them have been built according to the standards of science and with due consideration of empirical data. There is also a serious methodological problem that is, as a rule, overlooked: the need to take into consideration all of the empirical data and to reconstruct the Tunguska event before building any models of it. Such a reconstruction is essential – since the consequences of this event are many and varied. The main Tunguska traces may be grouped and listed as follows: (a) material traces; (b) instrumental traces; (c) informational traces. To be sure that a proposed theory is correct, the scientist must check it against all the three types of Tunguska evidence. Having reconstructed the Tunguska event with due attention to all the evidence, we have to conclude that it could not have been an asteroid or a comet core. There seems to exist in space another type of dangerous space objects, whose nature still remains unknown.

## I. Introduction

In the morning of June 30, 1908, a fiery body flew over the wastes of Central Siberia. It was clearly seen by inhabitants of the settlements situated on the banks of the Angara, Yenissey and Lena rivers, as well as by Tungus nomads in the taiga. The body's motion through the atmosphere was accompanied by thunderous sounds.

The strange object from space ended its flight path in a powerful explosion over the so-called Southern swamp, a small morass not far from the Padkamennaya Tunguska river. The coordinates of this location are: 60° 53′N & 101° 54′E. This airburst devastated about 2,150 km$^2$ of the taiga, flattening some 30 million trees. Over an area of 200 km$^2$ vegetation was burnt, which seems to be indicative of a powerful flash of light. (For details see: Vasilyev 2004; Rubtsov 2009; Rubtsov 2012.)

Some years later, the object that had exploded in June 1908 in Siberia was designated the "Tunguska meteorite." Whether or not this was a meteorite in the strict sense of this word remains unknown. It would therefore be more correct to call it the "Tunguska space body" (TSB). The moment of the Tunguska explosion has been determined with an accuracy of 10 sec. It occurred at 0 h 13 min 35 sec (± 5 sec) GMT (Pasechnik 1986:66). The accuracy of determination of the altitude of the explosion is not so good, but it is generally agreed that it was in the range from 6 to 8 km. But as for the total energy released at Tunguska, here the discrepancy between various estimations reaches more than two orders of magnitude: from $1.5\times10^{16}$ J (Boslough and Crawford 2008) to $2.9\times10^{18}$ J (Turco *et al*. 1982).

The main hypotheses proposed since 1927 to explain the Tunguska event can be listed as follows:

1. It was the arrival of a huge iron meteorite that broke into pieces high above the Earth's surface. Its large pieces and "a fiery jet of burning-hot gases" struck the surface and leveled the trees (Kulik 1927).
2. The impact of a huge iron or stony meteorite (Krinov 1949).
3. The forest devastation in the Tunguska taiga was caused by the bow wave which accompanied the meteorite flying in the atmosphere and hit the ground after the meteorite had been disrupted by the forces of air resistance (Rodionov and Tsikulin 1959).

4. Thermal explosion of the icy core of a comet (Krinov 1960).
5. A lump of “space snow” of extremely low density that completely collapsed in the atmosphere. Its bow wave leveled the taiga (Petrov and Stulov 1975).
6. The fast fragmentation of a stony asteroid or a comet core (Grigoryan 1976).
7. Low-altitude airburst of a swiftly moving stony asteroid (Boslough and Crawford 2008).
8. Vapor cloud explosion of a comet core (Tsynbal and Schnitke 1986).
9. Chemical explosion of a fragment of Comet Encke that was caught by the gravitational field of the Earth and made three revolutions around it, after which it entered the atmosphere and evaporated, forming an explosive cloud over Tunguska. Then the cloud detonated (Nikolsky, Schultz, and Medvedev 2008).
10. Annihilation of a meteorite consisting of antimatter (La Paz 1948).
11. Natural thermonuclear explosion of a comet core (D’Alessio and Harms 1989).
12. Nuclear explosion of an alien spacecraft (Kazantsev 1946).

Each of these hypotheses meets with considerable difficulties when trying to account for all of the peculiarities of this phenomenon, and therefore we do not possess as yet the correct theory.

## II. Methodology of the Investigation

The primary problem with the conventional interpretation of the Tunguska event is that there is no trace of either asteroidal or cometary material at the site of the explosion. Usually, authors of Tunguska hypotheses pay careful attention to this fact and try to build a mechanism explaining it, with varying degrees of success. But there is also a serious methodological problem that is, as a rule, overlooked: the need to take into consideration **all** of the empirical data and **to reconstruct the Tunguska event** before building any models of it. Such a reconstruction is essential – since the consequences of this event are many and varied. Meanwhile, more often than not, it is only some general characteristics of the leveled forest area that are taken into consideration when trying to find an explanation for the Tunguska event. There are, however, other traces of this event that should not be ignored. The main Tunguska traces may be grouped and listed as follows:

A. Material traces.
B. Instrumental traces.
C. Informational traces.

To be sure that a proposed theory is correct, we should check it against all the three types of Tunguska evidence.

## A. Material Traces

(1) **The leveling of trees over a butterfly-like area 70 km across and 55 km long** – the so-called "Fast's butterfly" (Fast 1967). Over this area, trees are lying mainly in a radial direction, although there are some slight deviations from this pattern. Its axis of symmetry runs at an angle of 115° to the east from its geographical meridian (Fig. 1, line A-B). Along this line the lying trees demonstrate a feeble herring-bone pattern apparently reflecting the action of the bow wave of the TSB on the forest. Because a bow wave travels symmetrically relative to the flying body's trajectory, this axis is in fact the projection of the trajectory. It attests that at the final stage of its flight the TSB was flying over the area of forest destruction in just this direction – that is, from the east-south-east to the west-north-west. The true azimuth of its flight direction was 295° (Fast 1967:60).

In the middle of the 1970s W. G. Fast and his colleagues, having studied additional data on the leveled forest collected in the field, concluded that there existed another belt of fallen trees showing a feeble herring-bone pattern and running at an angle of 99° to the east from its geographical meridian, that is practically from east to west (Fig. 1, line C-D). The true azimuth of the TSB's flight direction would therefore have been 279° (Fast, Barannik, and Razin 1976:48.) At the same time, Fast did not repudiate his earlier result. Hence, the pattern of forest destruction at Tunguska is quite complicated, suggestive of the effects of both a blast wave and **two** bow waves.

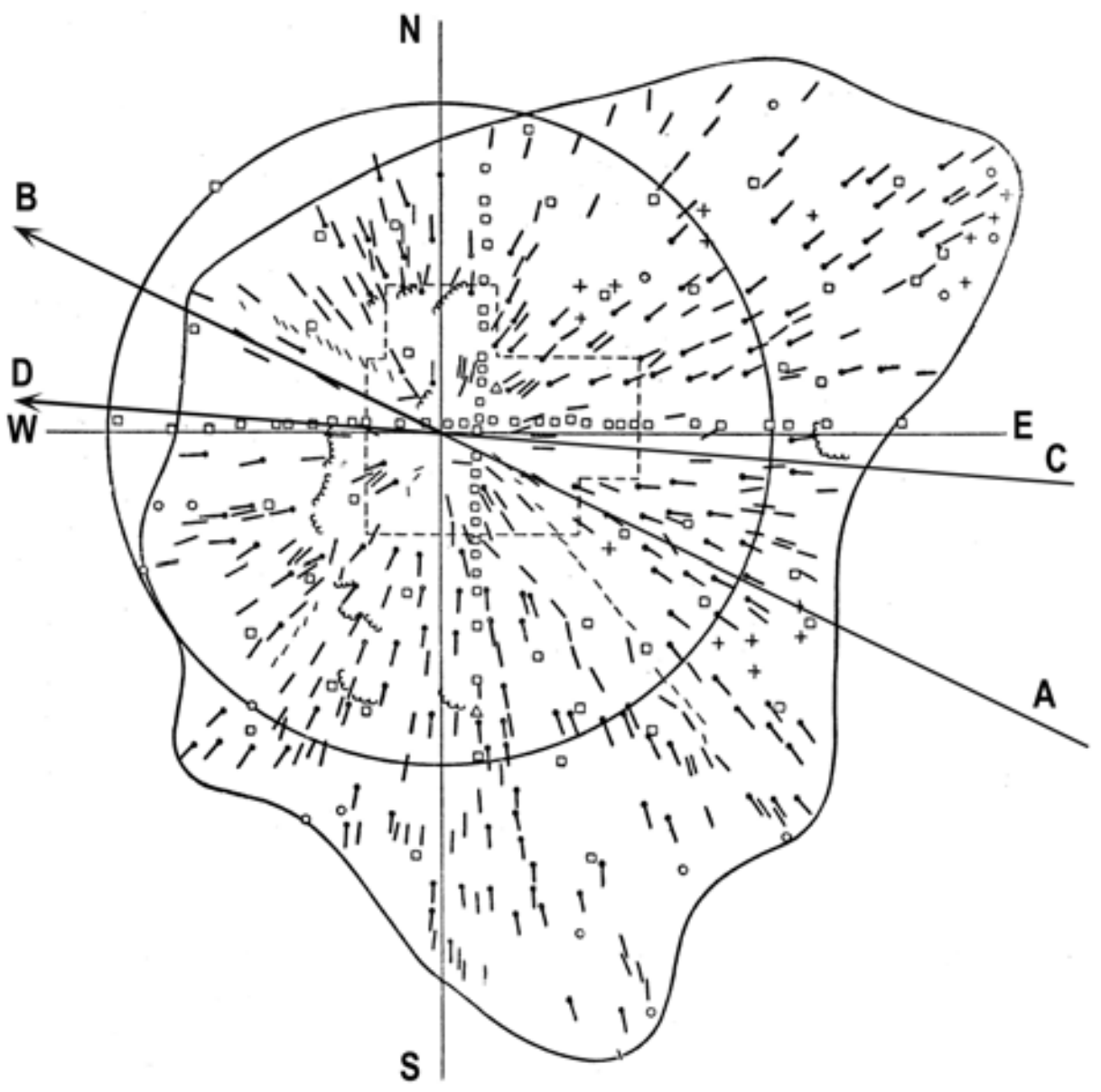

FIG. 1. The “Fast’s butterfly”: the outlines of the leveled forest at Tunguska, 2150 km$^2$ in size. Lines A-B and C-D designate the first and second TSB trajectories determined by W. G. Fast. *Based on*: (Boyarkina *et al*. 1964:127).

Quite remarkably, there is an area of about 8 km in diameter, at the epicenter of the explosion, where trees were scorched and devoid of branches, but remained standing upright like telegraph poles (Kulik 1927). The “telegraph-pole” phenomenon points to the effect of a blast wave, with its origin at a height of 6 to 8 kilometers. Although the Tunguska explosion is often referred to as “the largest impact event in recorded history,” there was in fact no impact in the strict sense of this word, that is no collision of a small space body with Earth’s surface.

Also, a trace of the bow wave in the leveled forest extends westward beyond the epicentral zone, which can mean that a fairly massive body flew westward **after** the explosion (Plekhanov and Plekhanova 1998).

(2) **The zone of the light burn of trees** (Fig. 2) is also “butterfly-like” in shape, its axis of symmetry running from the east to the west. It extends up to 16 km to the east from the epicenter, with two separate zones being noticeable within it: the zone of intense burns and the zone of weak burns.  Theoretically, traces of severe burning must have remained at the center of this figure and those of weak burning at its periphery. In reality the picture looks more complicated: the zone of weak burning extends from the east into the zone of severe burning; and along the axis of symmetry the burning is considerably weaker than that at a distance from it. At the very center of the figure there is evidence of the maximum level of the light flash.

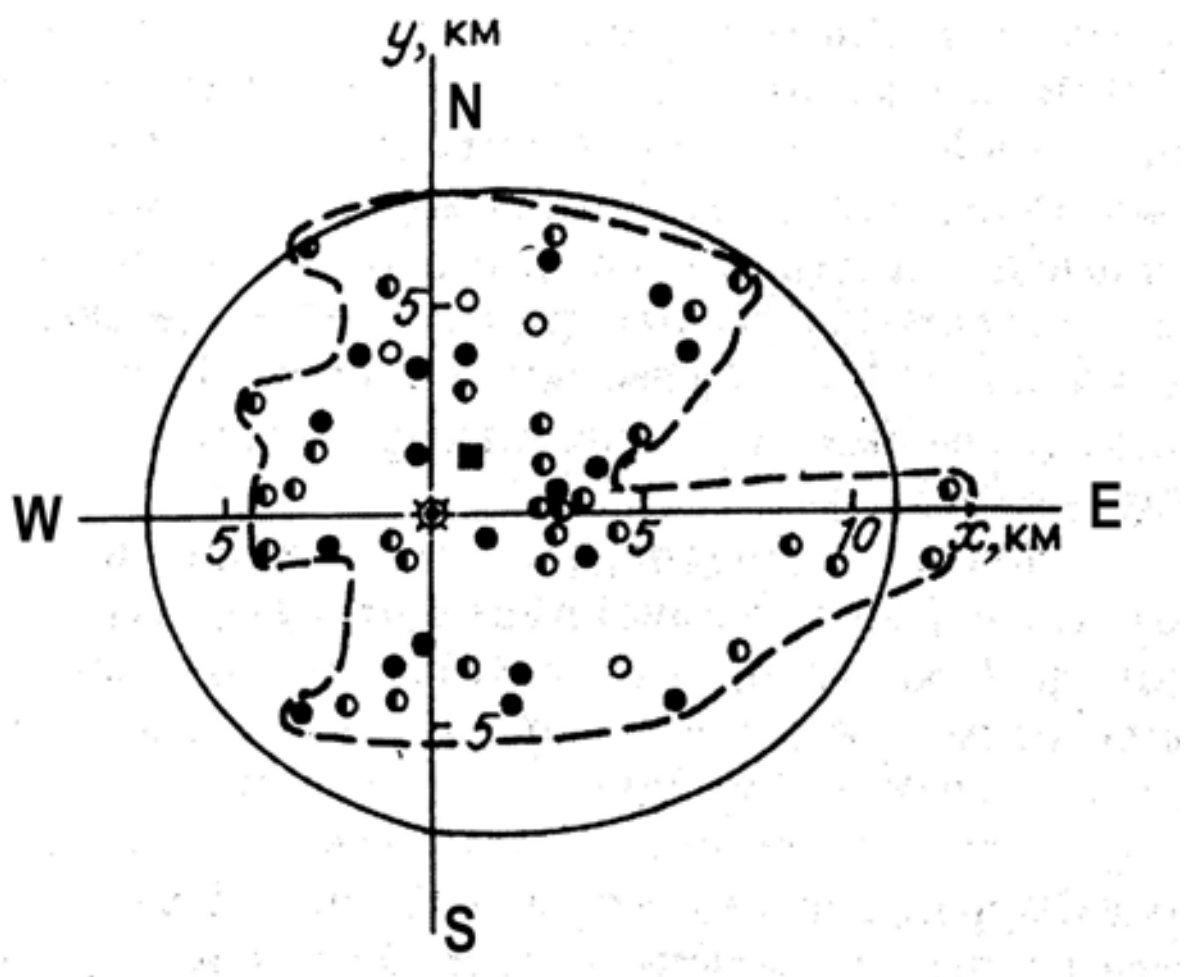

FIG. 2. Outlines of the Tunguska burned area from the light flash. *Source*: (Zhuravlev and Zigel 1998:103).

Having examined the traces of burning, V. K. Zhuravlev (1967) calculated that the heat radiation from the light flash, in the overall radiation of energy from the explosion, was not less than 10% and perhaps even 25% of the total energy released. It was therefore not only a high-altitude explosion but also a high-temperature one. Of course, for a high-temperature explosion of a comet core or stone asteroid the small cosmic body must have moved through the atmosphere at a great velocity (tens of kilometers per second).

(3) Are there any **material remnants of the TSB substance** at Tunguska? Although some silicate and metallic (containing cosmochemical elements – nickel, iron and cobalt) spherules some 100 μ in diameter were discovered in Tunguska peat and soil, the number of these spherules is much too small even for an icy comet core, to say nothing about a stony asteroid. The overall mass of space matter spread over Tunguska in 1908 was a few tons at best (Vasilyev 1986:6). But a powerful explosion of the comet core entering the Earth's atmosphere could have happened only if both its mass and velocity had been very high. According to well-justified estimations, the mass of the hypothetical Tunguska comet could not have been less than $10^6$ T (Fesenkov and Krinov 1960:35), perhaps even $10^7$ T (Tsynbal and Schnitke 1986:102). Most probably these microscopic spherules were due to the usual background fall of extraterrestrial matter.

However, some local geochemical anomalies have been discovered at the epicenter of the Tunguska explosion. The soil and peat are enriched with rare earths (samarium, europium, terbium, ytterbium, yttrium, etc), as well as with barium, mercury, copper, titanium, zinc and some other elements (Golenetsky and Stepanok 1980:113; Dmitriev and Zhuravlev 1984:34). Furthermore, the ratio of rare earth elements is anomalous. The content of terbium exceeds the norm by 55 times, that of thulium by 130 times, that of europium by 150 times, and that of ytterbium by 800 times (Dozmorov 1999).

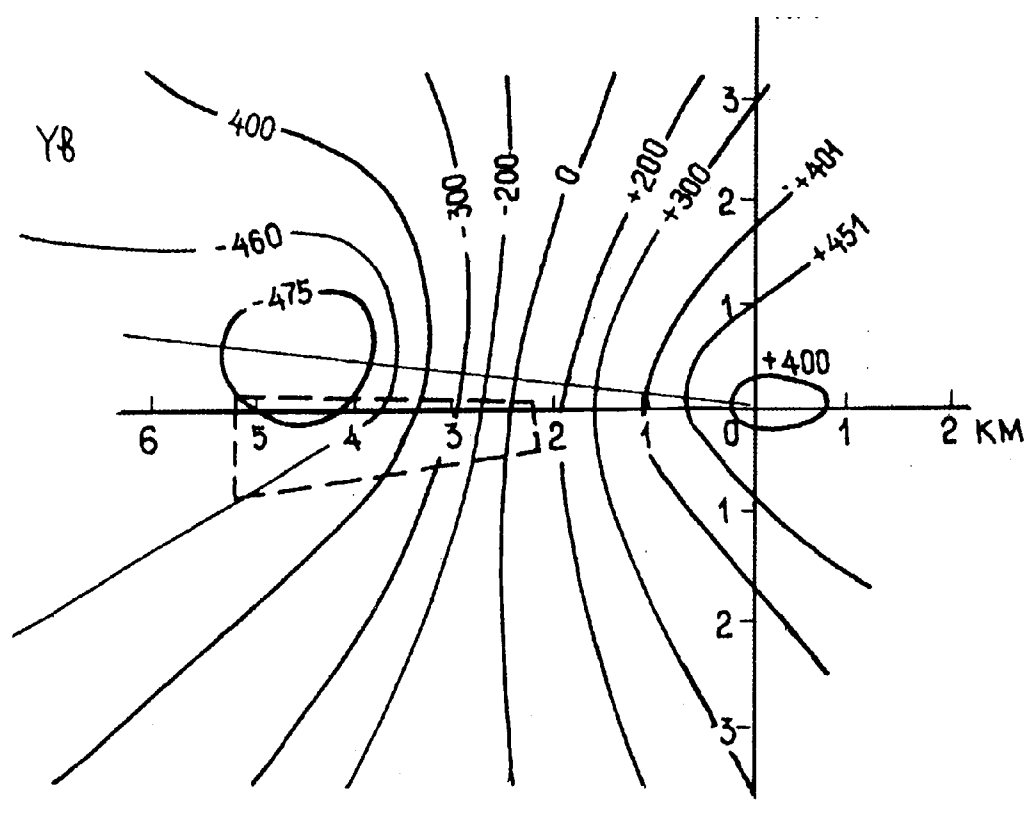

FIG. 3. Pattern of ytterbium's distribution at Tunguska following the projection of the first TSB trajectory determined by W. G. Fast. *Source:* (Zhuravlev and Demin, 1976:101).

Patterns of similar shapes are found at Tunguska for the surface distributions of lanthanum, lead, silver and manganese, but for iron, nickel, cobalt and chromium, the patterns of their distribution had no association with any special points or directions of the area of leveled forest, indicating that these elements were natural components of the soil and rocks. This can mean that usual cosmochemical elements – iron, nickel, cobalt – have nothing to do with the Tunguska space body. Instead, it is primarily ytterbium which can be reliably associated with the TSB.

Also, possibly lanthanum, lead, silver and manganese (Zhuravlev and Demin 1976:102). With this composition, it could hardly have been an asteroid or a comet core.

(4) **A complex set of serious ecological consequences** has been revealed in the region of the explosion. These are: first, a very fast restoration of the forest after the catastrophe and accelerated growth of trees, both young and those which survived the incident (Nekrasov and Emelyanov 1963; Emelyanov *et al*. 1967); and second, a sharply increased frequency of mutations in the local pines (Plekhanov *et al*. 1968; Dragavtsev *et al*. 1975). There was also discovered a rare mutation among the natives of the region, which arose in the 1910s in one of the settlements nearest to the epicenter (Rychkov 2000).

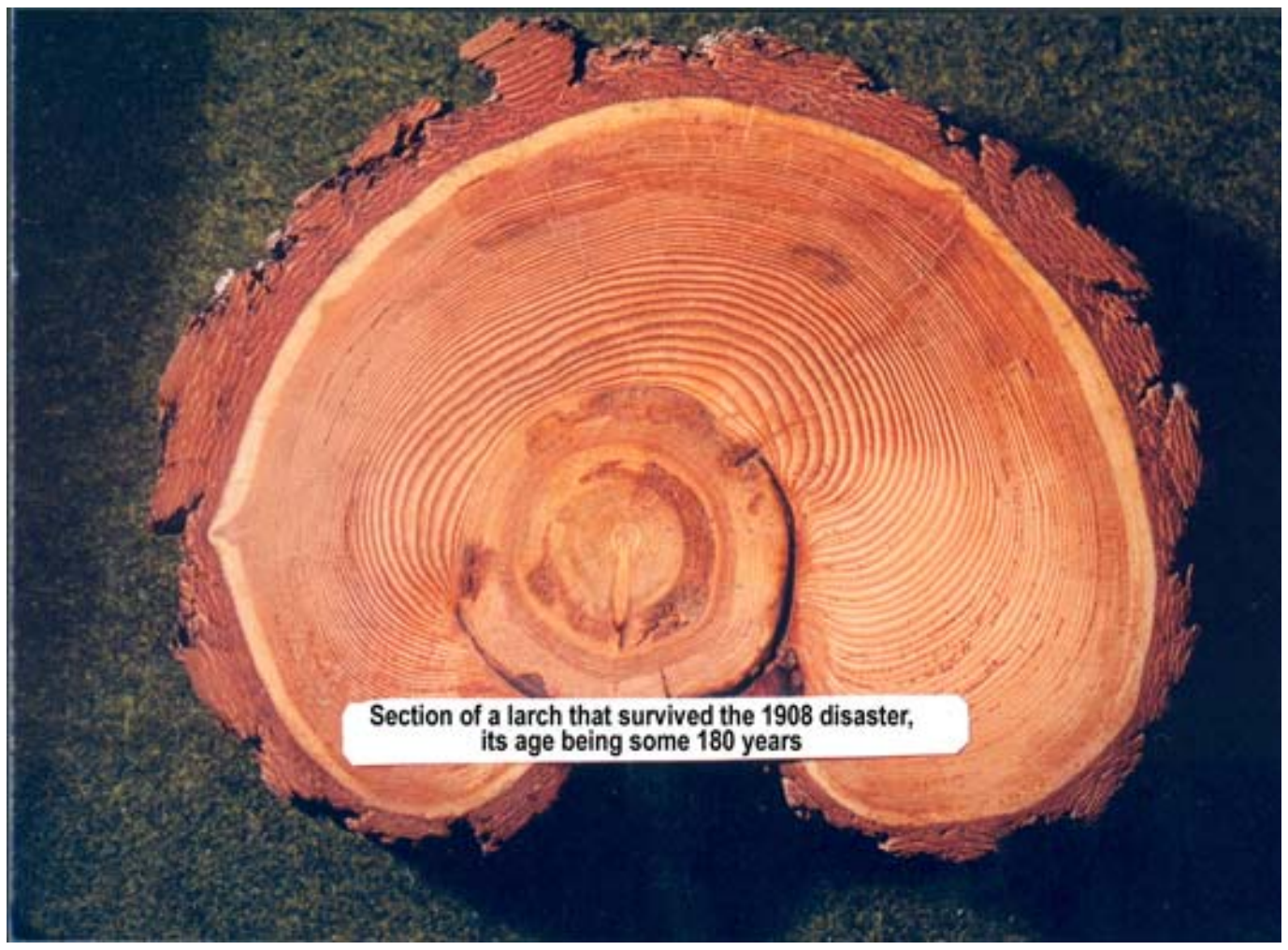

FIG. 4. A section of a larch that survived the 1908 disaster. Its rings after 1908 are noticeably wider than before. *Credit*: Vitaly Romeyko, Moscow, Russia.

(5) The presence of feeble but noticeable **radioactive fallout** after the Tunguska explosion is an empirical fact, confirmed by finding the peaks of radioactivity dated 1908 in trees that had withered before 1945 (that is, before the year when nuclear tests in the atmosphere started and the artificial radionuclides began to fall from the sky in abundance). Only the increased radioactivity of the samples taken from the trees that continued their growth after this year may be explained as contamination from contemporary nuclear tests (Mekhedov 1967;

Zolotov 1969). Note that the problem of Tunguska radioactivity was studied not by amateurs, but by the most distinguished Russian radiochemists, in particular by Professor Boris Kurchatov, the father of Soviet radiochemistry, and his close associate Dr. Vladimir Mekhedov (see: Vasilyev and Andreev 2006).

(6) Within 10 to 15 kilometers from the Tunguska epicenter the level of thermoluminescence (TL) of local minerals considerably exceeds the background level. The zone of the increased TL level has an axis of symmetry running almost directly from the east to the west. “Formerly we were calling the factor which had stimulated thermoluminescence at Tunguska somewhat too cautiously ‘unknown,’ but now it’s time to tell that we cannot see any rational alternatives to identifying this with hard radiation” (Bidyukov 2008:83).

The traces 4, 5 and 6 seem to indicate that the Tunguska explosion was accompanied by hard radiation.

## B. Instrumental Traces

(7) The Tunguska explosion left **records of its seismic waves** on the bands of seismographs in Irkutsk, Tashkent, Tbilisi and Jena.

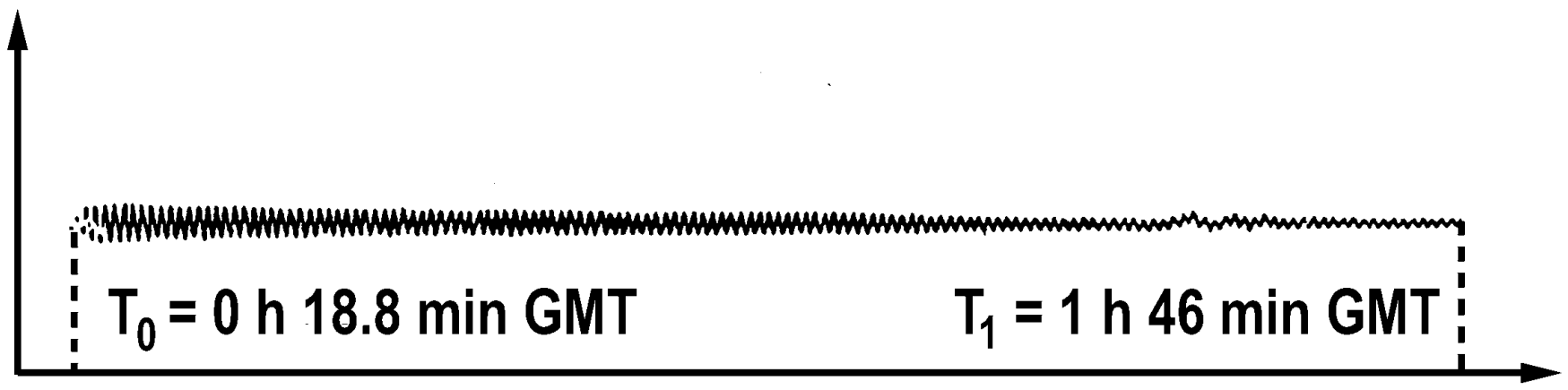

FIG. 5. The seismogram of the Tunguska earthquake of June 30, 1908 recorded by a seismograph of the Irkutsk Magnetographic and Meteorological Observatory. *Source:* (Krinov 1949:73).

In 1976, the leading Russian specialist in monitoring foreign nuclear tests Professor I. P. Pasechnik was asked by the Committee on Meteorites of the USSR Academy of Sciences to determine the magnitude of the Tunguska explosion. He examined in great details these seismograms and concluded that the most probable magnitude of the explosion was 30 to 50 megatons (Pasechnik 1976:51).

(8) **Microbarographs in Russia and in Britain have recorded infrasonic waves** of the explosion at Tunguska.

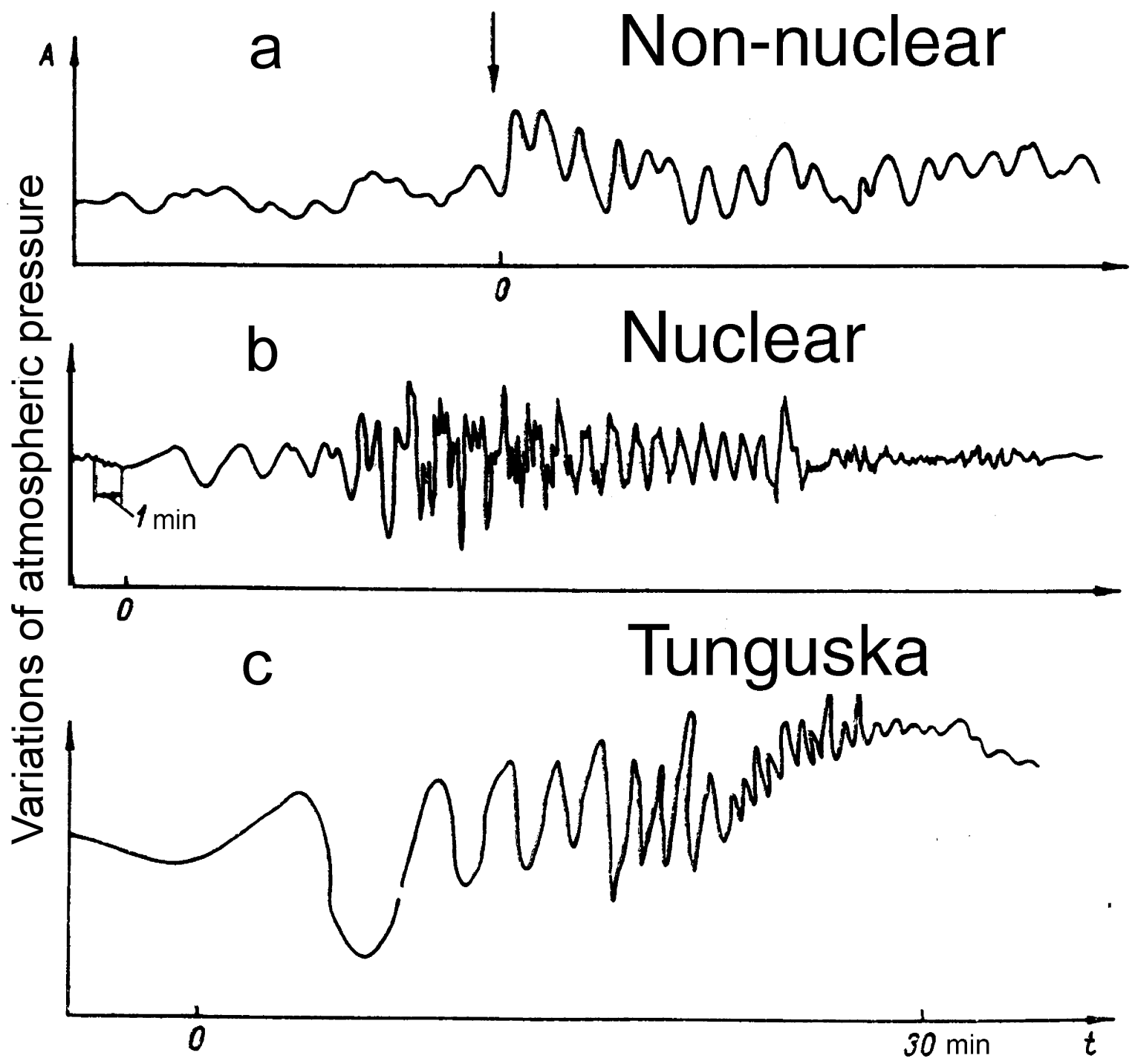

FIG. 6. Comparison of microbarograms of a nuclear, non-nuclear and Tunguska explosions. *Source*: (Zolotov 1969:150).

Attempting to explain the Tunguska explosion, authors of various hypotheses have used almost all known types of explosions: physical (impact, thermal, and dynamical, such as the swift fragmentation of the meteor body); chemical, including the vapor cloud explosion; and nuclear (fusion, fission, and antimatter annihilation). It is known that the nuclear explosion differs from all other types of explosion by its much greater concentration of energy (around $8.4\times10^{10}$ J/g, compared with $4.2\times10^{3}$ J/g). Thus, according to the concentration of energy all explosions may be separated into two groups: nuclear (having a high concentration of energy) and non-nuclear (having a low concentration of energy).

The "signatures" of nuclear and non-nuclear explosions on microbarograms are radically different. The most evident difference between them lies in the shape of the curve. The microbarogram of an explosion having a low ("non-nuclear") concentration of energy looks like a wave whose amplitude and period remain practically constant. For an explosion with a high ("nuclear") concentration of energy the curve on the tape of a microbarograph will be different: the amplitude and the period of this wave swiftly diminish with time. It is thanks to these characteristics of air waves that specialists monitoring nuclear tests can say immediately, not awaiting for information about nuclear contamination of the atmosphere, whether a powerful explosion detected by their instruments at a distant region of our planet was nuclear or not (Pasechnik 1962).

In Fig. 6, (*a*) microbarograms of a powerful chemical explosion are represented; and (*b*) a nuclear explosion with magnitude of several megatons that was carried out at a US testing ground on Marshall Islands in 1954. A third curve (*c*) is a record of air waves from the Tunguska explosion. The recording was made in 1908 in London. One can see that the curve *c* is very similar to the curve *b*, but has no resemblance to the curve *a*. Hence, A. V. Zolotov concluded that "the explosion of the Tunguska space body had a very high concentration of energy in a small volume" (Zolotov 1967:2094). Later, Pasechnik confirmed Zolotov's conclusion about the high concentration of the energy of the Tunguska explosion (Pasechnik 1976:51).

(9) Minutes after the explosion **a magnetic storm** began, that lasted some five hours. This storm was detected only by the Magnetographic and Meteorological Observatory in Irkutsk. No other magnetometric station on this planet had detected it (Ivanov 1964:144).

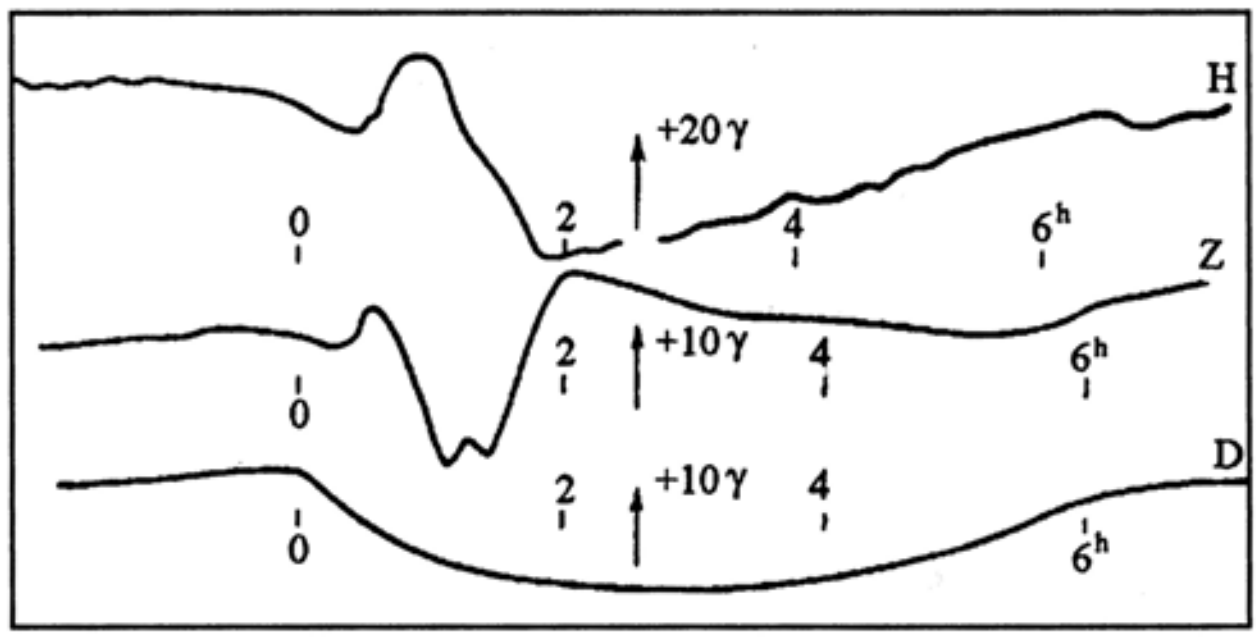

FIG. 7. The local geomagnetic storm, dated June 30, 1908, as recorded by instruments of the Magnetographic and Meteorological Observatory at Irkutsk. *Source:* (Ivanov 1961).

During 7 hours before the explosion of the Tunguska space body, the geomagnetic field was very calm. At 0 h 20 min GMT, that is 6 min after this body exploded, the intensity of the geomagnetic field abruptly increased by 4 nT and remained at that level for about 2 min. This was the initial phase of the local geomagnetic storm (called the “first entry”). Then began a second phase – “the rise phase.” In the course of 18 min it rose 20 nT more. The geomagnetic field reached its maximum intensity at 0 h 40 min GMT, and remained at the same level for the next 14 min. It then began to drop, the amplitude decreasing by some 70 nT. It returned to its initial undisturbed level only 5 hours later. Such effects have never been observed by astronomers studying meteor phenomena – neither before nor after the Tunguska event. The only parallel for this was the artificial geomagnetic storms that occurred during the high-altitude nuclear tests (Ivanov 1964:145; Zhuravlev 1998:9).

The separate stages of such storms lasted 10 to 20 min, and the intensities of the geomagnetic field reached 50 nT. These local geomagnetic storms were first recorded in August 1958, when thermonuclear charges of some 4 Mt in magnitude were detonated over Johnston Island at altitudes of 76 and 42 km (Matsushita 1959; Mason and Vitousek 1959). As it was soon established, this effect was generated by hard radiation from the fiery ball of the high-altitude nuclear explosion (Leypunsky 1960). Under the influence of this radiation, the level of ionization of the ionosphere increases sharply, there appear in it electric currents, and a magnetic disturbance occurs.

Great pains have been taken to explain the Tunguska geomagnetic storm, while not referring to the nuclear model of this event – particularly, via the action of the blast wave or the bow wave from the flying TSB on the ionosphere. None of these attempts were successful (Zhuravlev 1998). The proposed non-nuclear mechanisms were especially ineffective when trying to explain the long duration of the Tunguska geomagnetic storm and the fact that it was a very local effect. In 2003, speaking in Moscow at “The 95th Anniversary of the Tunguska Problem” conference, K. G. Ivanov agreed that the blast wave in itself could not have produced the geomagnetic effect. Additional ionization of the ionosphere over the place of the explosion was necessary (Ivanov 2003).

## C. Informational Traces

(10) Certainly, material and instrumental traces are the primary ones. But **Tunguska eyewitness reports** should not be ignored either. "If we are trying to unveil the real Tunguska mystery, and not just solve an abstract mathematical problem, we must reject those solutions which are inconsistent with observational data" (Bronshten 1980:161). These reports can be considered as boundary conditions for the "Tunguska theories". If a theoretical model goes beyond these boundaries this means it has nothing to do with the real Tunguska phenomenon.

The total number of eyewitness testimonies is about 700 (Vasilyev *et al.* 1981). The TSB was seen at a distance of up to 1000 km from the place of its explosion. There are two main areas of eyewitness reports:

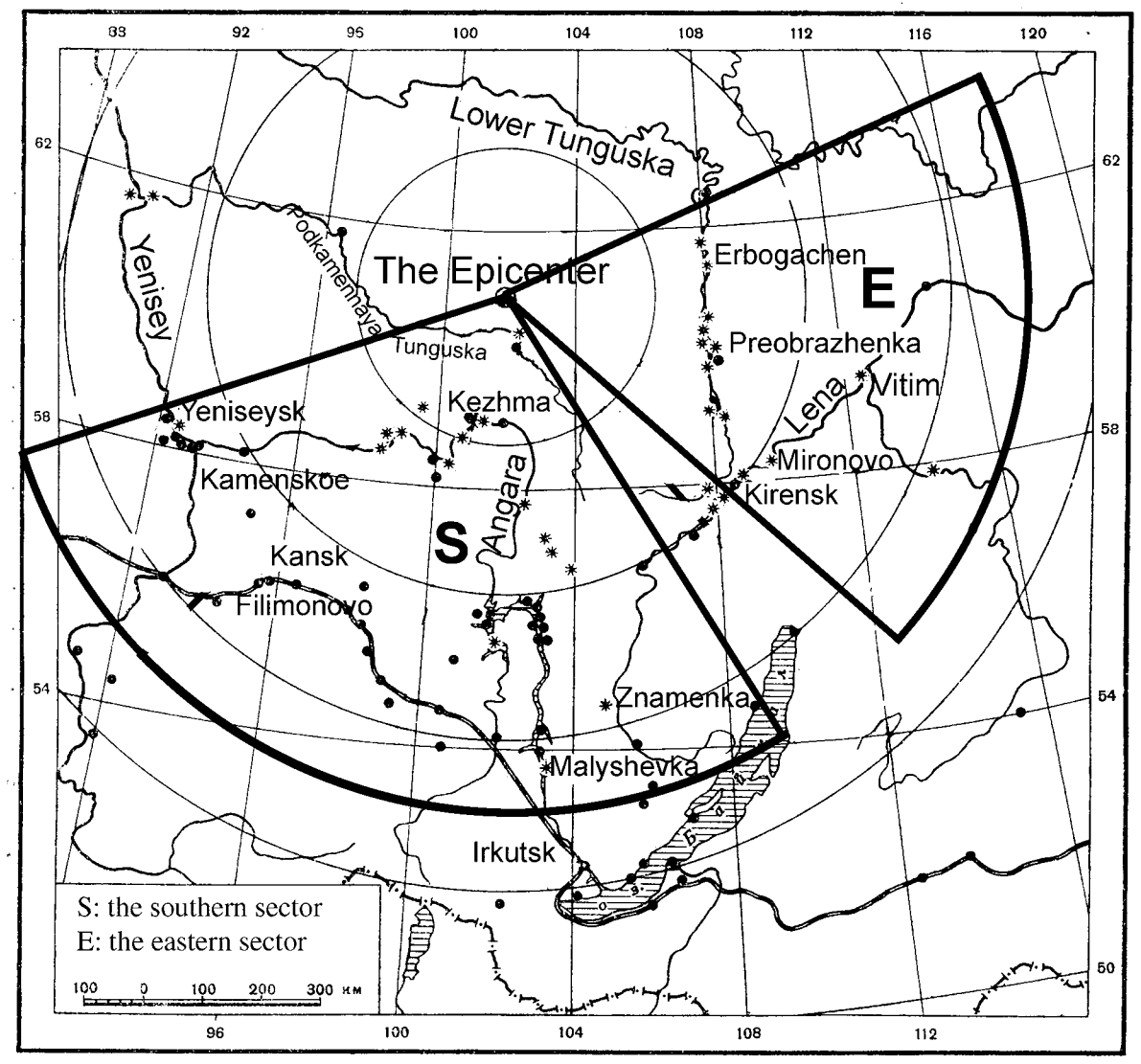

FIG. 8. The southern (S) and eastern (E) sectors, from which came reports of eyewitnesses observing the flight of the Tunguska space body (Rubtsov 2012:221).

Data obtained inside each sector made it possible to create a statistically reliable and coherent image of the Tunguska phenomenon, but these two images are different. In the south the phenomenon (including thunder-like sounds) lasted half an hour and more. The brightness of the TSB was comparable to the Sun. The body looked white or bluish. It had a short tail of the same color. After its flight there remained in the sky iridescent bands resembling a rainbow and stretching along the trajectory of the body's motion. And it flew from south to north.

In the east the brightness of the flying body was much lower than the Sun. Its color was red, and the shape was that of a ball or "artillery shell" with a long tail. Usually eyewitnesses said simply: a "red fiery broom" or a "red sheaf" was flying, and it was swiftly moving in the western direction, leaving no trace behind. The duration of this phenomenon did not exceed a few minutes.

The general scenario of the Tunguska event accepted as self-evident by the majority of Tunguska investigators is very simple: one space body flew over Central Siberia, generating in its flight a bow wave and performing no maneuvers, exploding over the Southern swamp and producing a blast wave. But when we process the eyewitness reports, we obtain, instead of an unambiguous picture of a space body arriving from a definite direction, either two bodies flying in different trajectories or one body performing various maneuvers – or a combination of these. Furthermore, if the TSB was seen at a distance of 1000 kilometers from the epicenter it means that it was flying at a small angle with respect to the Earth's surface. This angle could not have exceeded 10 to 15 degrees, otherwise the altitude at which a comet core or an asteroid began to emit light would have been too great. But in this case, the speed of the TSB before its explosion (that is, near the Southern swamp) could not have exceeded 1 to 2 km/sec, otherwise the body, flying in a flat trajectory, would have left in the leveled forest a more pronounced trace of its bow wave than it did leave (Rubtsov 2012:275). At this velocity no explosion due purely to the kinetic energy of a moving body is conceivable. So the TSB's explosion must have been produced by the internal energy of its substance (chemical, nuclear, or other).

## IV. Reconstruction of the Tunguska Event

It seems conceivable that in the morning of June 30, 1908, two space objects (let's call them TSB-A and TSB-B) flew over Central Siberia and one of them (TSB-A) exploded at Tunguska due to its internal energy, its concentration approaching that of a nuclear explosion. The explosion was accompanied by ionizing radiation and radioactive fallout. The ionizing radiation induced a magnetic disturbance in the ionosphere, which developed into a local geomagnetic storm lasting about five hours. The TSB-B had somehow survived this fiery bath and had flown farther west.

This is a somewhat simplified reconstruction of the Tunguska event – although revealing its most essential features. For a more detailed reconstruction see: (Rubtsov 2012:272-288).

## V. Conclusion

One must admit that the reconstructed image of the Tunguska phenomenon does not offer a definite answer to the question “What was it?” What is more, none of the existing hypotheses fits this image sufficiently well. In particular, the high concentration of energy of the Tunguska explosion contradicts the hypothesis of the vapor cloud explosion. And an ordinary comet or a stony asteroid seems to be out of the question.

It goes without saying that the Tunguska “meteorite” (TSB-A) was a dangerous space body of unknown nature. Had it exploded over London or New York an entire city would have been destroyed. What kind of body was it? At present we do not know. But we know that instead of elements that are prevalent in space – iron, nickel and cobalt, it contained titanium, aluminum, ytterbium and gold... The Tunguska space body flew at a low velocity and exploded due to the internal energy of its substance, not due to the energy of motion. Its explosion had a high concentration of energy, approaching that of a nuclear explosion. Also, it was accompanied by ionizing radiation and radioactive fallout.

So, if the TSB was a natural space body, then it means that there exists in space another type of dangerous space objects, whose nature remains vague at best. Naturally enough, to estimate chances of their collision with our planet and predict their coming, it will be needed, first of all, to detect these space bodies instrumentally and to determine their physical properties and parameters of their orbits. Until then, the only thing we can say about these objects is that they are very different from asteroids and comets.